# Anomalous Hall effect and spin orbit torques in MnGa/IrMn films: Modification from strong spin Hall effect of antiferromagnet

K. K. Meng [1, *, †], J. Miao[1], X. G. Xu[1], Y. Wu[1], X. P. Zhao[2], J. H. Zhao[2] and Y. Jiang[1, *, ‡]

[1] *School of Materials Science and Engineering, University of Science and Technology Beijing, Beijing 100083, China*

[2] *State Key Laboratory of Superlattices and Microstructures, Institute of Semiconductors, Chinese Academy of Sciences, Beijing 100083, China*

**Abstract:** We report systematic measurements of anomalous Hall effect (AHE) and spin orbit torques (SOT) in MnGa/IrMn films, in which a single $L1_0$-MnGa epitaxial layer reveals obvious orbital two-channel Kondo (2CK) effect. As increasing the thickness of the antiferromagnet IrMn, the strong spin Hall effect (SHE) has gradually suppressed the orbital 2CK effect and modified the AHE of MnGa. A scaling involving multiple competing scattering mechanisms has been used to distinguish different contributions to the modified AHE. Finally, the sizeable SOT in the MnGa/IrMn films induced by the strong SHE of IrMn have been investigated. The IrMn layer also supplies an in-plane exchange bias field and enables nearly field-free magnetization reversal.

**Keywords:** Spin polarized transport in metals; Spin–orbit coupling; Spin transport effects

**PACS:** 72.25.Ba; 71.70.Ej; 75.76.+j

[*]Authors to whom correspondence should be addressed:

[†] kkmeng@ustb.edu.cn

[‡] yjiang@ustb.edu.cn

## I. INTRODUCTION

Spin Hall effect (SHE) that converts charge currents into spin currents in a heavy metal (HM) with strong spin-orbit coupling (SOC) has attracted much interest due to its practical use in technological applications [1-8]. When a HM comes in contact with a ferromagnet (FM), the spin currents that diffuse into the FM will modify the spin dependent transport properties such as anomalous Hall effect (AHE) [9]. On the other hand, the spin currents will also exert torque on the magnetization, which is similar to spin transfer torque [10-16]. The investigation and clarification of these intriguing issues are of fundamental importance for better understanding the underlying physics.

The AHE is the most prominent phenomenon that exists in FM, which originates from the interplay between SOC and magnetism [17]. Electrons moving through a FM will acquire a transverse velocity with opposite directions for different spin orientations due to SOC, since the charge currents have usually a net polarization, this spin-dependent transverse velocity will result in a net transverse anomalous Hall voltage [18]. Despite the discovery of AHE more than a century ago, only recently a more universal microscopic description is emerging. Karplus and Luttinger have proposed that the intrinsic AHE arises from the transverse velocity of Bloch electrons induced by SOC together with interband mixing [19-24]. Smit and Berger suggested the extrinsic mechanisms including skew scattering and side jump coming from the asymmetrical scattering of conduction electrons due to SOC [25-27]. To explore and distinguish the possible mechanisms of AHE, a unified scaling describing the AHE resistivity $\rho_{AH}$ in terms of the longitudinal resistivity $\rho_{xx}$ has always been investigated [28-33]. Considering the essence of AHE, the modification from SHE can also be explored through clarifying $\rho_{AH} \sim \rho_{xx}$ relations [9].

Another spin dependent transport behavior is Kondo effect, which is a striking consequence of conduction electrons coupling with localized spin or "pseudo-spin" impurities. The orbital two-channel Kondo (2CK) effect displaying exotic non-Fermi liquid (NFL) behavior arises in the intricate scenario of two conduction electrons compensating a pseudo-spin 1/2 impurity of two-level system (TLS) [34]. The orbital 2CK effect from TLSs is manifested in electrical transport by a unique temperature ($T$) dependence in resistivity with three distinct $T$ regimes: a low-$T$ upturn characterized by $\Delta\rho_{xx} \sim \ln T$ for $T$ beyond Kondo temperature $T_K$, followed by NFL behavior $\Delta\rho_{xx} \sim T^{1/2}$ for $T_D < T < T_K$ and deviation from $T^{1/2}$ dependence upon further cooling [34, 35]. The $T^{1/2}$ dependence is a hallmark of the NFL state in the orbital 2CK effect, in striking contrast to the $T^2$ scaling of Fermi-liquid behavior in fully screened Kondo effect. In a conventional FM, although the Kondo coupling between the TLS and itinerant electrons is irrelevant to the electron spins, the symmetry of the two spin channels is broken due to ferromagnetic exchange splitting of the $d$-band. The channel asymmetry should lead to different tunneling rates of a TLS for the two spin channels and thus weaken the NFL behavior. If the channel asymmetry is large enough, it should be manifested as a decreased magnitude and an enhanced effective breakdown temperature of the 2CK effect as a result of the enhanced decoupling of TLS from one spin channel [34]. Therefore, the orbital 2CK effect in the FM will be suppressed when contacting with a strong SOC metal, since the pure spin current generated by SHE will diffuse into the FM layer and enhance the channel asymmetry.

Moreover, in HM/FM heterostructures with perpendicular magnetic anisotropy (PMA), the generated pure spin current can also exert torques on the magnetization, which is referred as to spin orbit torques (SOT) [19-24]. On the other hand, spin accumulation can take place at the FM/HM interface via the Rashba effect, which has also generated significant effective torques and caused current-induced domain nucleation and fast domain wall motion. In most previous experiments on SOT, an extra in-plane magnetic field is indeed required to achieve deterministic switching, and this is detrimental for device applications. More recently, however, the experimental demonstrations of strong SOC in antiferromagnets (AFMs) provide an alternative to heavy metals in HM/FM herostructures for devices with active SOT [13, 36, 37]. By replacing heavy metals with AFMs, the purely electrical deterministic switching of perpendicular magnetization without any assistance from an external magnetic field will be realized, since an AFM can supply an exchange bias field that can serve as an effective magnetic field.

In this context, as an attempt to offer a step towards exploring these rich and underlying physics, we have systematically investigated the AHE and SOT in MnGa/IrMn films, in which a single $L1_0$-MnGa epitaxial layer reveals obvious orbital 2CK effect. The strong SHE of the AFM IrMn has suppressed the orbital 2CK effect and modified the AHE of MnGa. The results are compared with MnGa/Al and MnGa/Pt films. A scaling involving multiple competing scattering mechanisms has been used to distinguish different contributions to the modified AHE. It is found that the spin-dependent transport properties in MnGa/HM (IrMn or Pt) are much different from that in the MnGa single layer and the MnGa/Al films, in which Al displays weak SHE. On the other hand, the sizeable SOT in the MnGa/IrMn films induced by the strong SHE of IrMn has also been observed. The IrMn layer supplies an in-plane exchange bias field and enables nearly field-free magnetization reversal. By performing adiabatic harmonic Hall voltage measurements, we have also quantitatively investigated the SHE induced effective field.

## II. EXPERIMENTAL DETAILS

A 3-nm-thick $L1_0$-MnGa single-crystalline film was grown on a semi-insulating GaAs (001) substrate by molecular-beam epitaxy at substrate temperature $T_S=250$ °C. Here the Mn/Ga atoms ratio is 1 [38]. Then, IrMn films with different thickness were immediately deposited on it by dc magnetron sputtering. Finally, a 1.5-nm-thick Pt layer was deposited for preventing oxidation. After deposition, to provide the exchange bias, the samples were annealed at 250 °C for 30 minutes under an in-plane magnetic field of 5 KOe along the +**X** direction. In the experiments, the 6-nm-thick Al and Pt were also deposited on MnGa for contrast experiments. Photolithography and Ar ion milling were used to pattern Hall bars and a lift-off process was used to form the contact electrodes. A scanning electron microscope (SEM) image of a patterned Hall bar and the schematic of the measurement setup along with the definition of the coordinate system used in this study are shown in Figure 1(a). The size of all the Hall bars is 10 μm×80 μm. Two electrodes for current injection are labelled **I**$_+$ and **I**$_-$. Another two electrodes for the Hall voltage measurements are labelled **V**$_+$ and **V**$_-$. We measured the SOT induced magnetization switching by applying a pulsed current with the width 50 μs, and the resistance was measured after a 16 μs delay under an external magnetic field $\mathbf{H_X}$ along either positive or negative $\mathbf{X}$ directions. We apply a sinusoidal AC current with the frequency of 259.68 Hz to exert periodic SOT on the magnetization. The first $V_\omega$ and the second $V_{2\omega}$ harmonic anomalous Hall voltages were measured as a function of the magnetic field $\mathbf{H}$ at the same time using two lock-in amplifier systems.

## III. RESULTS AND DISCUSSION

The saturated magnetization $\mathbf{M_S}$ of MnGa is 170 emu/cc, which is much smaller than the theoretical value for the fully ordered $L1_0$-MnGa probably due to the loss of the chemical ordering. The effective anisotropy fields $\mathbf{H_K}$ is given by $\mathbf{H_K}=\mathbf{H_{sat}}+4\pi\mathbf{M_S}$, where $\mathbf{H_{sat}}$ is the hard-axis saturation field. For MnGa/IrMn($t$), $\mathbf{H_K}$ is calculated to be about 1.3 T, 1.1 T and 0.8 T respectively as increasing $t$ from 4 to 8 nm. The normalized anomalous Hall resistance $\mathbf{R_{AH}}$ versus the perpendicular magnetic field for MnGa/IrMn ($t$=0, 4, 6, 8 nm) were measured with applying a DC current of 1 mA as shown in Fig. 1(b). It was obtained by subtracting the ordinary Hall component determined from a linear fit to the high-field region up to ±6T. As $t$ increases, the squareness of the loop decreases due to the increasing magnetic coupling between MnGa and IrMn. The inset figure shows the AHE resistivity $\rho_{AH}$, which decreases as increasing $t$ from 4 to 8 nm. We have also measured the temperature dependent resistivities of single Pt, Al and IrMn films. Then, assuming that each film in the MnGa/Pt, MnGa/Al and MnGa/IrMn bilayers acts as a parallel resistance path, both the AHE resistivity $\rho_{AH}$ and the longitudinal resistivity $\rho_{xx}$ in this manuscript have been expressed as those of the MnGa layer.

Fig. 2 (a) and (b) plot the $\mathbf{T}$ dependence of the resistivity variation at $\mathbf{H}$=0 Oe for the $L1_0$-MnGa film, which shows distinct signatures associated with the TLS-induced 2CK effect. The longitudinal resistivity $\rho_{XX}$ firstly varies linearly with $\ln\mathbf{T}$ as increasing $\mathbf{T}$ from 300 K. Then, $\rho_{XX}$ deviates from the ln $\mathbf{T}$ dependence and crossover to a $\mathbf{T}^{1/2}$ dependence when $\mathbf{T}$ drops below $\mathbf{T_K}$=115 K. The large $\mathbf{T_K}$ of the $L1_0$-MnGa film suggests a strong Kondo coupling between the TLS and conduction electrons. The $\mathbf{T}^{1/2}$-dependent resistivity is regarded as a unique signature of the NFL behavior for the 2CK effect. As further decreasing temperature, $\rho_{XX}$ begins to increase faster than $T^{1/2}$ below a characteristic temperature $T_D$=45 K, indicating the deviation from the NFL behavior. This represents an observation of the TLS theory-expected deviation from the orbital 2CK state below $T_D$ in a diffusive conductor. After depositing a 6-nm-thick Al film, the orbital 2CK effect has not been changed evidently, as shown in Fig 2 (a) and (b). Notably, comparing the $T$ dependent of $\rho_{XX}$ in MnGa and MnGa/Al, evident suppression has been found in MnGa/IrMn (2 nm) as shown in Fig. 1(d) and (e). $\rho_{XX}$ for the MnGa/IrMn (2 nm) bilayer shows a resistivity minimum at a characteristic temperature ($T_M$=165 K). In the high $T$ regime ($T>T_M$), $\rho_{XX}$ increases linearly with $T$ due to the increasing phonon scattering, and $T_K$ decreases to 85 K. The suppression gradually increased as increasing the thickness of IrMn with further decreasing $T_K$ and $T_D$ as shown in Fig. 2(g)-(i). The 6-nm-thick Pt layer has also suppressed the orbital 2CK effect similar with MnGa/IrMn (8nm) as shown in Fig. 2(j). The values of $T_K$ and $T_D$ have been summarized in Table I. In a conventional ferromagnet, although the Kondo coupling between the TLS and itinerant electrons is irrelevant to the electron spins, the symmetry of the two spin channels is broken due to ferromagnetic exchange splitting of the $d$-band. The channel asymmetry should lead to different tunnelling rates of the TLS for two spin channels and thus weaken the NFL behaviors in comparison with its nonmagnetic counterpart ($\Delta N=N_\uparrow-N_\downarrow=0$), where $N_\uparrow$ and $N_\downarrow$ are the density of states for majority and minority spin channels in the conduction band. Based on the Stoner model for itinerant ferromagnetism, $M_S$ can be used as an index for the degree of spin population imbalance of the conduction band. The small $M_S$ of the MnGa film studied here indicates a very low degree of spin population imbalance with the loss of the chemical ordering. This could be the reason why the ferromagnetic exchange splitting does not quench the 2CK physics. Considering the enhanced channel asymmetry will weaken the NFL behavior, we suppose that the orbital 2CK effect will be suppressed if the pure spin current generated by the metals with strong SOC, such as Pt or IrMn, diffuses into the FM. The Fermi surface will be modified and the enhanced $\Delta N$ will quench the orbital 2CK effect. To establish more rigorously the orbital 2CK effect in our $L1_0$-MnGa films, we have examined the effect of the applied perpendicular magnetic fields, $\mathbf{H_Z}$, on the $\mathbf{T}$-dependent resistivity as shown in the Fig. 3. The $\mathbf{H}$-independent resistivity upturn scaling with ln $\mathbf{T}$ and $\mathbf{T}^{1/2}$ in the two $\mathbf{T}$ regimes, and the deviation from the NFL behavior at the lowest temperatures are consistent with the TLS model.

For clarity we normalize the resistivity of all the samples by the value at 300 K and plot the temperature dependence of $\rho_{XX}$ $(\mathbf{T})/\rho_{XX}$ (300 K) in Fig. 4(a), and the inset figure shows the $\rho_{XX}$ for all the six films. Fig. 4(b) shows the $\mathbf{T}$ profiles of $\rho_{AH}$ for all the six samples, which all decrease as cooling down. So far, there have been few reports on a similar anomaly in the $\mathbf{T}$ dependence of $\rho_{AH}$ [40]. We infer that the strong orbital 2CK effect in $L1_0$-MnGa could be the reason for such a complex $\mathbf{T}$ dependence of $\rho_{AH}$. We plot $\rho_{AH}$ as a function of $\rho_{xx}$ for all the samples to compare the difference of scaling curves in Fig. 4(c). Most recently, Hou *et al* have derived a general scaling form of the anomalous Hall resistivity as a function of multiple scattering mechanisms using [31]:

$$\rho_{AH} = \alpha \rho_{xx0} + \beta_0 \rho_{xx0}^2 + \gamma \rho_{xx0} \rho_{xxT} + \beta_1 \rho^2_{xxT} , \quad (1)$$

where $\rho_{xx0}$ is the residual resistivity induced by impurity scattering, $\rho_{xxT} = \rho_{xx} - \rho_{xx0}$ is due to dynamic disorders (mainly phonons at higher temperatures), $\alpha$ is the skew scattering coefficient, the coefficients $\beta_0$, $\beta_1$ and $\gamma$ contain intrinsic Berry curvature and side-jump contributions. Here $\rho_{xx0}$ is considered as the minimum resistivity. As shown in Fig. 5, the new scaling works well for all samples except for $t$=4 and 6 nm at low temperatures, where the orbital 2CK effect becomes important. The coefficients $\alpha$, $\beta_0$, $\gamma$, and $\beta_1$ determined from Eq. (2) for all the samples are shown in Table I. $\alpha$ is large negative in both MnGa and MnGa/Al (6 nm), which should be due to the extrinsic AHE contribution from impurities and static defects with the loss of chemical ordering. However, it changes into relatively small positive in MnGa/Pt(6 nm) and MnGa/IrMn($t$). Considering the chemical ordering of MnGa has not been changed, the variation of skew scattering should partly come from the asymmetric scattering due to the effective SOC at the interface and partly come from the modification of the Fermi surface with injected spin-polarized current. On the other hand, as discussed by Hou *et al*., it seems challenging to separate out the intrinsic part from the side-jump one in $\beta_0$, $\beta_1$ and $\gamma$. However, $(\beta_0+\beta_1-\gamma)$ and $(\gamma-2\beta_1)$ are the coefficients of pure scattering effect and they have also changed sign with contacting HMs. Totally speaking, the contacting HMs have introduced spin-dependent scattering at interface due to SOC and spin polarized current which modified the Fermi surface of MnGa.

To further verify the modification from SHE, we have carried out SOT measurements for MnGa/IrMn ($t$=4, 6, 8 nm) and MnGa/Pt(6 nm) at room temperature. The current-induced magnetization switching with applying an in-plane field of $\mathbf{H_X}$ is shown in Fig. 3. For MnGa/IrMn(4 nm), the magnetization is switched from +$\mathbf{Z}$ to -$\mathbf{Z}$ with $\mathbf{H_X}$=+3 KOe when sweeping the current from positive to negative, and switched back from -$\mathbf{Z}$ to +$\mathbf{Z}$ when sweeping the current reversely. With $\mathbf{H_X}$=-3 KOe, the opposite switching behavior is observed. The switching current density is about $1.5 \times 10^8$ A/cm$^2$. As decreasing the absolute value of $\mathbf{H_X}$, the magnetization could not been fully switched, and there is no switching at $\mathbf{H_X}$=0 Oe. Therefore, the SHE of IrMn does exert torques on the magnetization of MnGa, but there is nearly no exchange bias field. It also indicates that the IrMn has a positive spin Hall angle, which is consistent with a previous study on the inverse SHE in IrMn [40, 41]. As increasing $t$ to 6 nm, the switching behavior at positive and negative applied field is not symmetry. Partial magnetization has been switched at 0 Oe, indicating a weak exchange bias field has been induced in the films. This phenomenon becomes more evident in MnGa/IrMn(8 nm), in which a nearly field-free switching has been observed. In our experiments, the exchange bias effect happened along both parallel and perpendicular axes. Firstly, as shown in Fig. 1(b), as increasing the thickness of IrMn to 8 nm the coercivity of the bilayers has been decreased from 5 KOe to 1 KOe. On the other hand, according to the SOT results, the exchange bias effect also happened along $\mathbf{X}$ axis, and the exchange bias field is about 2~3 KOe. The AFM IrMn has not only generated SOT but also supplied an in-plane exchange bias field, which enables all-electrical deterministic switching of perpendicular magnetization in the absence of magnetic field. However, for MnGa/Pt (6 nm), it is found that the magnetization of MnGa cannot be fully switched, mostly because the spin Hall angle of Pt is smaller than IrMn and the SHE is weak.

To determine the strength of spin-orbit effective fields in the MnGa/IrMn films, we have performed non-resonant magnetization-tilting measurements by applying a small amplitude low frequency alternating current (AC) with the amplitude of 2.1 mA through the device and simultaneously sweeping a static in-plane magnetic field parallel or perpendicular to the current direction ($\mathbf{H_X}$ and $\mathbf{H_Y}$). The damping-like $\mathbf{H_D}$ and field-like effective fields $\mathbf{H_F}$ can be calculated by [6]

$$H_{D(F)} = -2\frac{\partial V_{2\omega}/\partial H_{X(Y)}}{\partial^2 V_{\omega}/\partial H_{X(Y)}^2}, \qquad (2)$$

Fig. 7 shows the first harmonic $V_{\omega}$ and second harmonic Hall voltages $V_{2\omega}$ plotted against the in-plane external field $\mathbf{H_X}$, which are measured with out-of-plane magnetization component $M_Z>0$ and with $M_Z<0$. However, there is no obvious $V_{2\omega}$ signal with $\mathbf{H_Y}$ applied, indicating that the field-like SOT is largely eliminated in the films. We have also extracted the effective spin Hall angle $\theta_{SH}=0.27$ of IrMn in MnGa/IrMn and $\theta_{SH}=0.11$ of Pt in MnGa/Pt using the relation $H_D = \hbar\theta_{SH}|J_C|/(2|e|M_S t_F)$, where $\mathbf{J_C}$ is charge current density, $e$ is the charge of an electron, $M_S$ is the saturation magnetization of MnGa, and $t_F$ is the thickness of the MnGa [42]. The value of IrMn is larger than the result in the previous studies [41], which may be ascribed to the different experimental method. Fig. 8(a) shows the estimated $\mathbf{H_D}$ with out-of-plane magnetization component $M_Z<0$ using Eq. (2) as applied current density $\mathbf{J_C}$. The effective field varys linearly with $\mathbf{J_C}$, indicating that the effects of Joule heating are negligible in the measured $\mathbf{J_C}$ range. The magnitude of $\mathbf{H_D}$ increases to be 30 Oe per $10^6$A/cm$^2$ as $t$ increases to be 8 nm. On the other hand, Anomalous Nernst effect (ANE) also has contribution to the second harmonic voltage. We have measured the ANE contribution in MnGa/IrMn(8nm) with field sweeping along the out-of-plane direction similar with Ref [16] as shown in Fig. 8 (b). By comparing $V_{ANE}$ and $V_{2w}$, one can see that ANE contribution is very small.

**IV. SUMMARY**

In conclusion, we have systematically investigated the AHE and SOT in the MnGa/IrMn films, in which a single $L1_0$-MnGa epitaxial layer reveals obvious orbital 2CK effect. The strong SHE of IrMn has suppressed the orbital 2CK effect and modified the AHE of MnGa. The results are compared with the MnGa/Al and MnGa/Pt films. A scaling involving multiple competing scattering mechanisms has been used to distinguish different contributions to the modified AHE. It is found that these two kinds of spin dependent transport properties in the MnGa/HM (IrMn or Pt) are quite different from those in the MnGa single layer and MnGa/Al films in which Al displays very weak SHE. On the other hand, the sizeable SOT in the MnGa/IrMn films induced by the strong SHE of IrMn has also been observed. The IrMn layer supplies an in-plane exchange bias field and enables nearly field-free magnetization reversal. By performing adiabatic harmonic Hall voltage measurements, we also quantitatively investigated the SHE induced effective field.

**ACKNOWLEDGMENTS**

This work was partially supported by the National Basic Research Program of China (2015CB921502), the National Science Foundation of China (Grant Nos. 61404125, 51371024, 51325101, 51271020).

**References**

[1] J. E. Hirsch, Phys. Rev. Lett. **83**, 1834 (1999).

[2] I. M. Miron, K. Garello, G. Gaudin, P. J. Zermatten, M. V. Costache, S. Auffret, S. Bandiera, B. Rodmacq, A. Schuhl, and P. Gambardella, Nature **476**, 189 (2011).

[3] L. Q. Liu, C.-F. Pai, Y. Li, H. W. Tseng, D. C. Ralph, and R. A. Buhrman, Science **336**, 555 (2012).

[4] K. Garello, I. M. Miron, C. O. Avci, F. Freimuth, Y. Mokrousov, S. Blügel, S. Auffret, O. Boulle, G. Gaudin, and P. Gambardella, Nat. Nanotechnol. **8**, 587 (2013).

[5] H. Nakayama, M. Althammer, Y.-T. Chen, K. Uchida, Y. Kajiwara, D. Kikuchi, T. Ohtani, S. Geprägs, M. Opel, S. Takahashi, R. Gross, G. E. W. Bauer, S. T. B. Goennenwein, and E. Saitoh, Phys. Rev. Lett. **110**, 206601 (2013).

[6] J. Kim, J. Sinha, M. Hayashi, M. Yamanouchi, S. Fukami, T. Suzuki, S. Mitani, and H. Ohno, Nat. Mater. **12**, 240 (2013).

[7] L. Q. Liu, O. J. Lee, T. J. Gudmundsen, D. C. Ralph, and R. A. Buhrman, Phys. Rev. Lett. **109**, 096602 (2012).

[8] S. Emori, U. Bauer, S.-M. Ahn, E. Martinez, and G. S. D. Beach, Nat. Mater. **12**, 611 (2013).

[9] K. K. Meng, J. Miao, X. G. Xu, J. X. Xiao, J. H. Zhao, and Y. Jiang, Phys. Rev. B **93**, 060406(R) (2016).

[10] C. O. Avci, K. Garello, C. Nistor, S. Godey, B. Ballesteros, A. Mugarza, A. Barla, M. Valvidares, E. Pellegrin, A. Ghosh, I. M. Miron, O. Boulle, S. Auffret, G. Gaudin, and P. Gambardella, Phys. Rev. B **89**, 214419 (2014).

[11] G. Yu, P. Upadhyaya, Y. Fan, J. G. Alzate, W. Jiang, K. L. Wong, S. Takei, S. A. Bender, L.-T. Chang, Y. Jiang, M. Lang, J. Tang, Y. Wang, Y. Tserkovnyak, P. K. Amiri, and K. L. Wang, Nat. Nanotechnol. **9**, 548 (2014).

[12] C. Zhang, S. Fukami, H. Sato, F. Matsukura, and H. Ohno, Appl. Phys. Lett. **107**, 012401 (2015).

[13] S. Fukami, C. L. Zhang, S. DuttaGupta, A. Kurenkov, and H. Ohno, Nat. Mater. **15**, 4 (2016).

[14] T. Suzuki, S. Fukami, N. Ishiwata, M. Yamanouchi, S. Ikeda, N. Kasai, and H. Ohno, Appl. Phys. Lett. **98**, 142505 (2011).

[15] X. Qiu, K. Narayanapillai, Y. Wu, P. Deorani, D.-H. Yang, W.-S. Noh, J.-H. Park, K.-J. Lee, H.-W. Lee, and H. Yang, Nat. Nanotechnol. **10**, 333 (2015).

[16] X. Qiu, P. Deorani, K. Narayanapillai, K.-S. Lee, K.-J. Lee, H.-W. Lee, and H. Yang, Sci. Rep. **4**, 4491 (2014).

[17] E. H. Hall, Philos. Mag. **10**, 301 (1880).

[18] A. Hoffmann, IEEE Trans. Magn. **49**, 5172 (2013).

[19] R. Karplus and J. M. Luttinger, Phys. Rev. **95**, 1154 (1954).

[20] G. Sundaram and Q. Niu, Phys. Rev. B **59**, 14915 (1999).

[21] T. Jungwirth, Q. Niu, and A. H. MacDonald, Phys. Rev. Lett. **88**, 207208 (2002).

[22] M. Onoda and N. Nagaosa, J. Phys. Soc. Jpn. **71**, 19 (2002).

[23] J. Sinova, D. Culcer, Q. Niu, N. A. Sinitsyn, T. Jungwirth, and A. H. MacDonald, Phys. Rev. Lett. **92**, 126603 (2004).

[24] D. Xiao, M.-C. Chang, and Q. Niu, Rev. Mod. Phys. **82**, 1959 (2010).

[25] J. Smit, Physica (Amsterdam) **21**, 877 (1955).

[26] L. Berger, Phys. Rev. B **2**, 4559 (1970).

[27] A. Crepieux and P. Bruno, Phys. Rev. B **64**, 014416 (2001).

[28] S. Onoda, N. Sugimoto, and N. Nagaosa, Phys. Rev. Lett. **97**, 126602 (2006).

[29] T. Miyasato, N. Abe, T. Fujii, A. Asamitsu, S. Onoda, Y. Onose, N. Nagaosa, and Y. Tokura, Phys. Rev. Lett. **99**, 086602 (2007).

[30] Y. Tian, L. Ye, and X. Jin, Phys. Rev. Lett. **103**, 087206 (2009).

[31] D. Z. Hou, G. Su, Y. Tian, X. F. Jin, S. Y. A. Yang, and Q. Niu, Phys. Rev. Lett. **114**, 217203 (2015).

[32] J. L. Xu, Y. F. Li, D. Z. Hou, L. Ye, and X. F. Jin, Appl. Phys. Lett. **102**, 162401 (2013).

[33] C. Zeng, Y. Yao, Q. Niu, and H. H. Weitering, Phys. Rev. Lett. **96**, 037204 (2006).

[34] L. J. Zhu, S. H. Nie, P. Xiong, P. Schlottmann, and J. H. Zhao, Nat. Commun. **7**, 10817 (2016).

[35] D. L. Cox and A. Zawadowki, Adv. Phys. **47**, 599 (1998).

[36] Y. -W. Oh, S. C. Baek, Y. M. Kim, H. Y. Lee, K. -D. Lee, C. -G. Yang, E. -S. Park, K. -S. Lee, K. -W. Kim, G. Go, J. -R. Jeong, B. -C. Min, H. -W. Lee, K. -J. Lee and B. -G. Par, Nat. Nanotechnol. (2016); 10.1038.

[37] A. van den Brink, G. Vermijs, A. Solignac, J. Koo, J. T. Kohlhepp, H. J. M. Swagten, and B. Koopmans, Nat. Commun. **7**, 10854 (2016).

[38] L. J. Zhu, D. Pan, S. H. Nie, J. Lu, and J. H. Zhao, Appl. Phys. Lett. **102**, 132403 (2013).

[39] R. Mathieu, A. Asamitsu, H. Yamada, K. S. Takahashi, M. Kawasaki, Z. Fang, N. Nagaosa, and Y. Tokura, Phys. Rev. Lett. **93**, 016602 (2004).

[40] W. Zhang, M. B. Jungfleisch, W. Jiang, J. E. Pearson, A. Hoffmann, F. Freimuth, and Y. Mokrousov, Phys. Rev. Lett. **113**, 196602 (2014).

[41] W. Zhang, M. B. Jungfleisch, F. Freimuth, W. Jiang, J. Sklenar, J. E. Pearson, J. B. Ketterson, Y. Mokrousov, and A. Hoffmann, Phys. Rev. B **92**, 144405 (2015).

[42] R. Ramaswamy, X. Qiu, T. Dutta, S. D. Pollard, and H. Yang, Appl. Phys. Lett. **108**, 202406 (2016).

**Figure Captions**

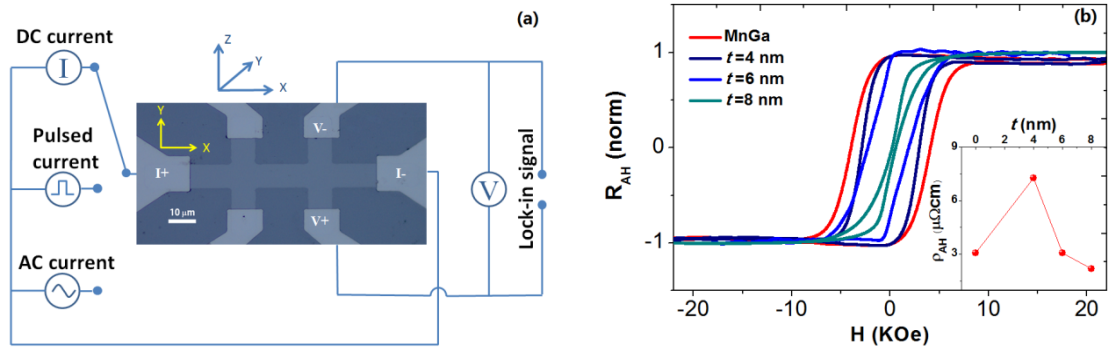

Fig. 1 (a) SEM image of a patterned Hall bar and the schematic measurement setup along with the definition of the coordinate system. (b) Out-of-plane hysteresis loop of the normalized anomalous Hall resistance in MnGa/IrMn($t$ nm). The inset shows the AHE resistivity $\rho_{\mathrm{AH}}$.

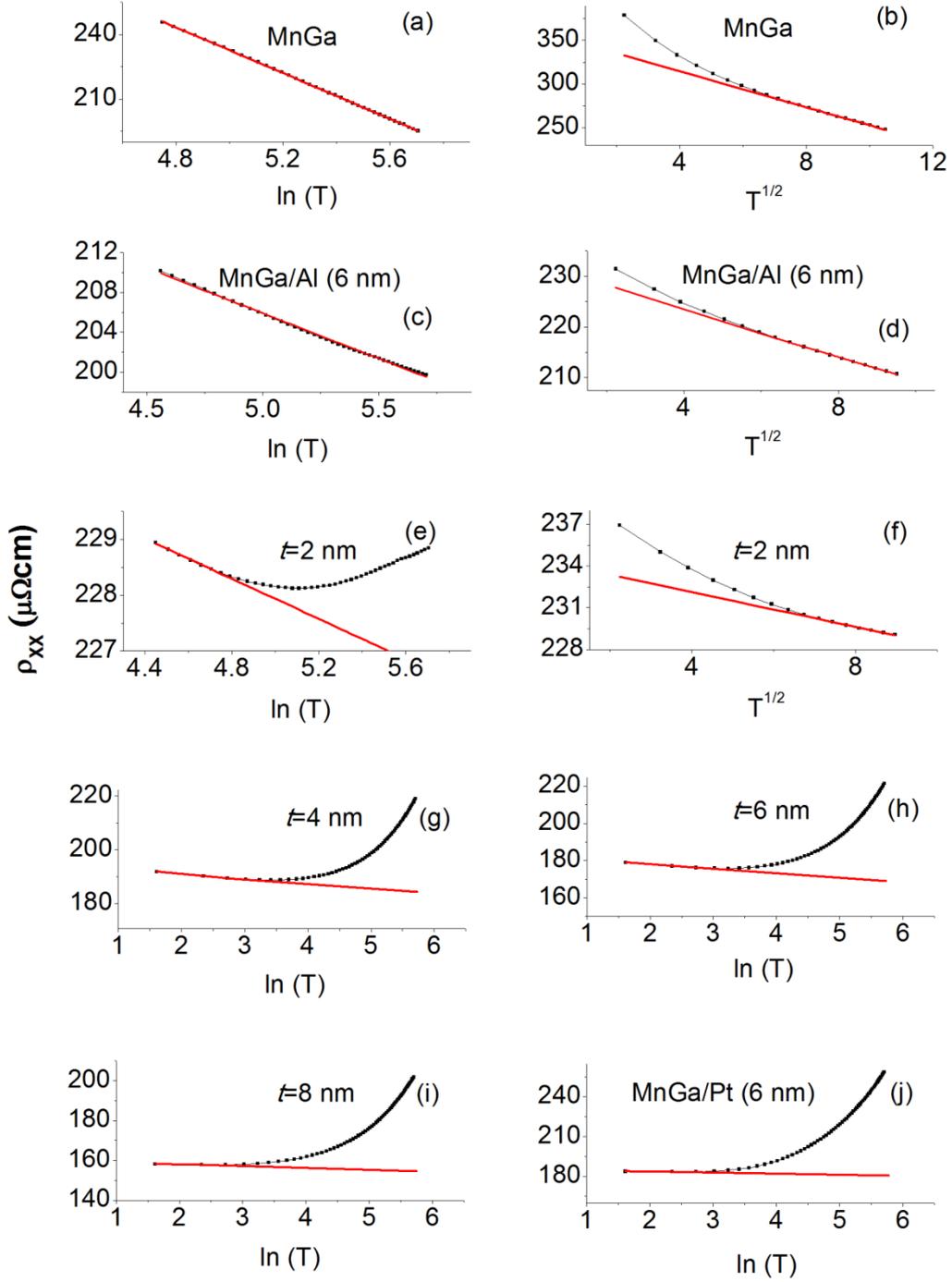

Fig. 2 The temperature dependence of the longitudinal resistivity $\rho_{xx}$ for all the samples.

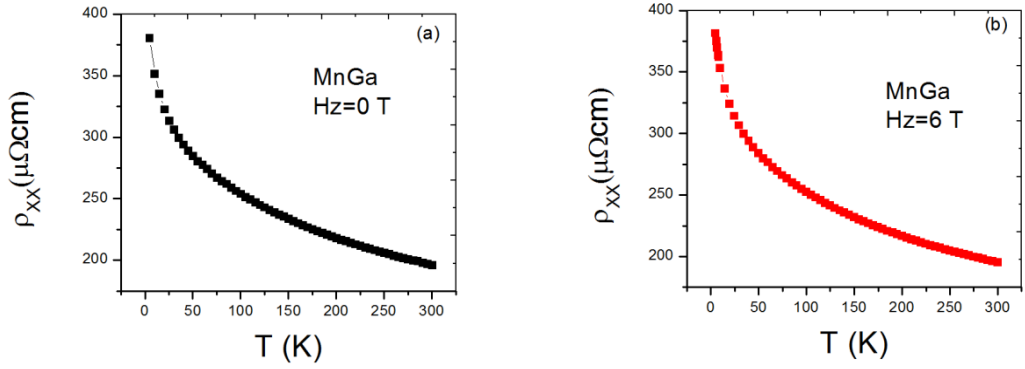

Fig. 3 $\rho_{xx}$ vs **T** at $\mathbf{H_Z}$=0 T (a) and $\mathbf{H_Z}$ =6 T (b) for the $L1_0$-MnGa single layer.

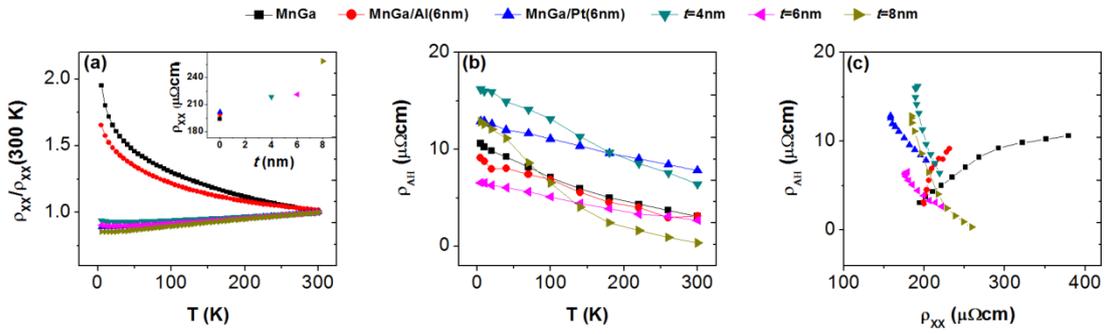

Fig. 4 (a) The temperature dependence of the longitudinal resistivity $\rho_{XX}$ for the six samples. The data are normalized by the values at 300 K. The inset shows $\rho_{XX}$ measured at 300 K. (b) The temperature dependence of $\rho_{AH}$ for the six samples. (c) $\rho_{AH}$ as the function of $\rho_{XX}$ for the six samples.

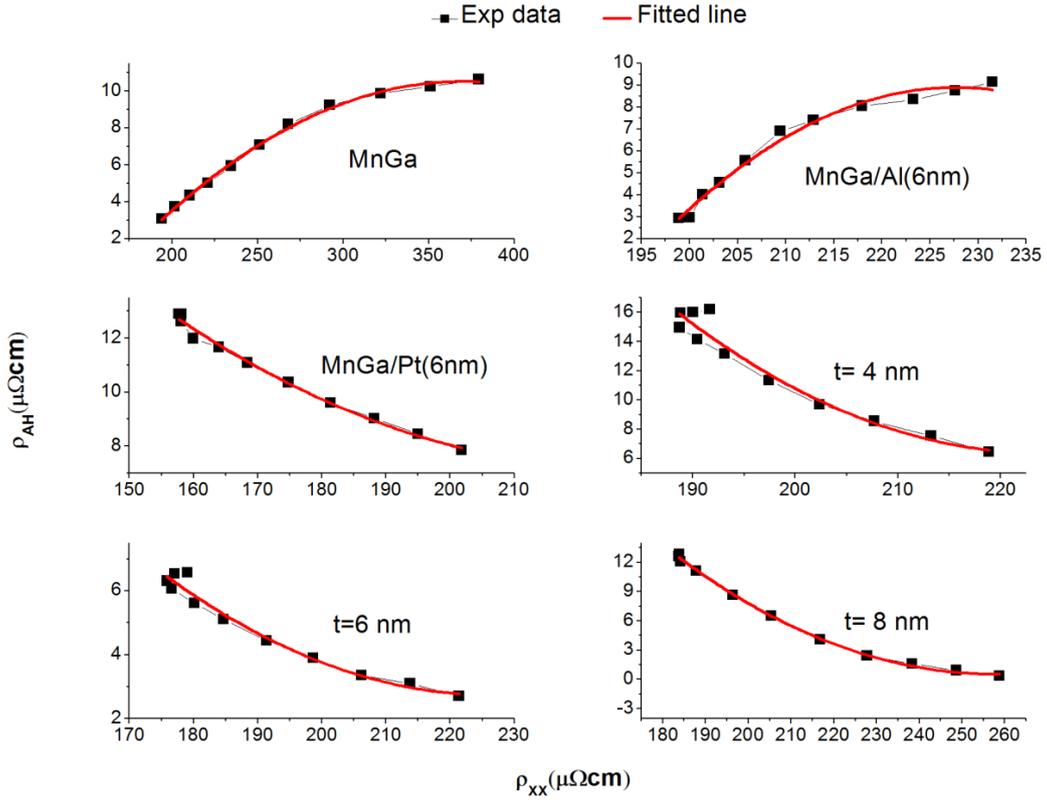

Fig. 5 $\rho_{AH}$ vs $\rho_{xx}$ for the six samples. Black squares are experimental raw data. The red lines are fitted plots using Eq. (1).

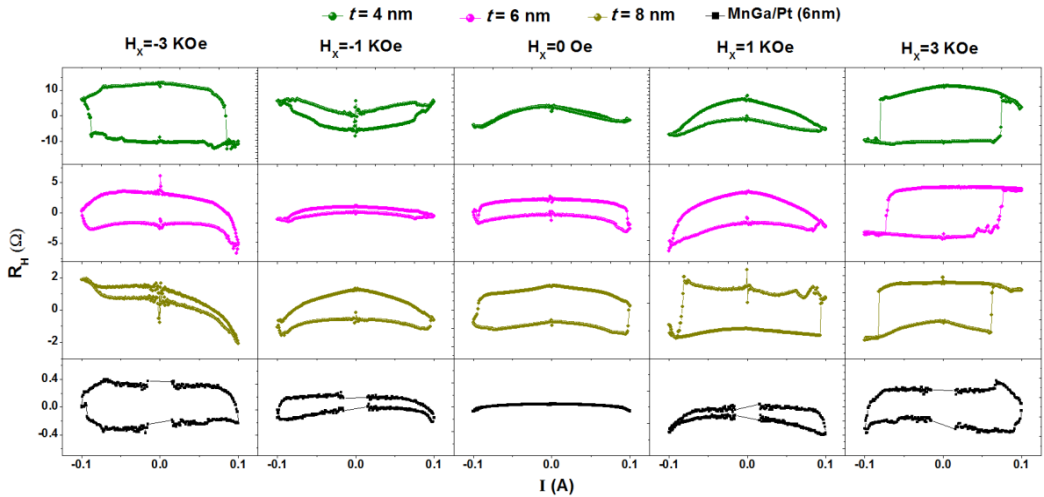

Fig. 6 **$R_H$-I** curves in MnGa/IrMn($t$=4, 6, 8nm) and MnGa/Pt (6 nm) under different in-plane external fields **$H_X$**.

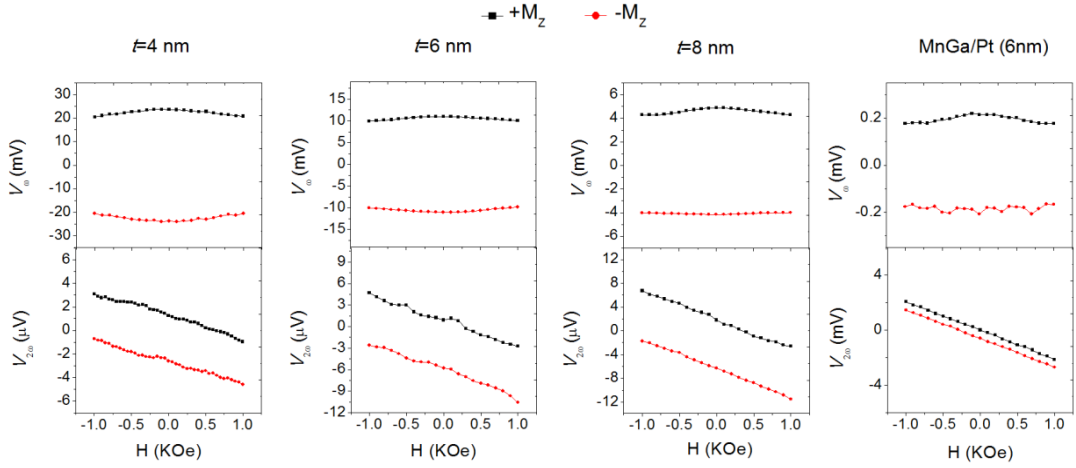

Fig. 7 The first $V_{\omega}$ and second $V_{2\omega}$ harmonic Hall voltages plotted against the in-plane external fields $\mathbf{H_X}$ for the four samples.

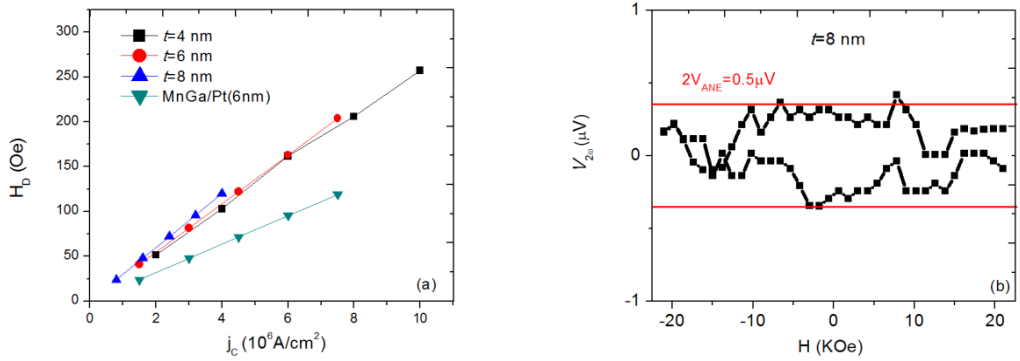

Fig. 8 (a) Current-induced effective field $H_D$ versus applied AC current density for the four samples. (b) Second $V_{2\omega}$ harmonic Hall voltages plotted against the out-of-plane external fields $\mathbf{H_Z}$ for MnGa/IrMn (8 nm).

TABLE I. $T_D$, $T_K$, the AHE scaling coefficients, current-induced effective field $H_D$ for the six samples.

| | $T_D$ (K) | $T_K$ (K) | $\alpha$ | $\beta_0$ ($\mu\Omega^{-1}\text{cm}^{-1}$) | $\gamma$ ($\mu\Omega^{-1}\text{cm}^{-1}$) | $\beta_1$ ($\mu\Omega^{-1}\text{cm}^{-1}$) | $H_D$ (Oe per $10^6\text{A/cm}^2$) |
|---|---|---|---|---|---|---|---|
| MnGa | 45 | 115 | -1.039 | -4.1e-4 | 4.38e-4 | -2.1e-4 | 0 |
| MnGa/Al(6) | 45 | 100 | -1.042 | -1.4e-3 | 6.4e-4 | -7.2e-3 | 0 |
| MnGa/Pt(6) | 45 | 85 | 0.249 | 1.2e-3 | -2.8e-4 | 1.1e-3 | 12 |
| $t$=4nm | ~ | 5 | 0.676 | 7.7e-3 | -4.3e-4 | 7.6e-3 | 27 |
| $t$=6nm | ~ | 5 | 0.524 | 1.4e-3 | -7.4e-4 | 1.4e-4 | 28 |
| $t$=8nm | ~ | 5 | 0.569 | 2.1e-3 | -1.5e-4 | 2.2e-3 | 30 |